# Nonperturbative isotope effect on light-matter interaction in boron arsenide

Huan Wu*
*School for Engineering of Matter, Transport and Energy, Arizona State University, Tempe, AZ 85287, USA*

The interaction of light and matter under strong isotope disorder gives rise to unconventional physics that goes beyond the quantum perturbation theory. In boron arsenide, the large mass difference between the two stable boron isotopes presents a paradigmatic case where perturbation theory fails, yet a unified theoretical framework across the perturbative and nonperturbative regime has remained elusive. Here, we develop a nonperturbative approach to capture isotope-disorder effect in boron arsenide, which fundamentally alters light-matter interactions. We reveal that coherently mixed vibrations between two boron isotopes reshape the dielectric function in the nonperturbative regime. The nonperturbative isotope interactions dictate the properties of coupled surface phonon polaritons and near-field radiative heat transfer. Two-fold tuning of radiative heat flux is achieved by modulating the surface phonon polariton resonance via isotope engineering. This work establishes a unified framework connecting the perturbative and nonperturbative limits, enabling quantitative predictions across weak to strong disorder regime.

## I. INTRODUCTION

Quantum perturbation theory (QPT), built on the weak-coupling assumption, has been the predominant approach for treating many-body interactions in condensed matter systems, such as electron-phonon, phonon-phonon, and light-matter interactions [1,2]. Within the QPT framework, the Hamiltonian $H$ is decomposed into a sum of a non-interacting component $H_0$, and a perturbative component $H_I$, such that $H = H_0 + H_I$. $H_0$ describes a simplified system whose eigenstates can be easily derived, whereas $H_I$ induces corrections such as energy shifts and linewidths of the eigenstates. With QPT, complicated systems can therefore be tackled based on the known solutions of simpler ones. While this simplified yet powerful framework effectively tackles weakly interacting regimes, it breaks down in the strong-interaction regime, where $H_I$ cannot be treated as a small perturbation. In such cases, nonperturbative approaches become essential for describing emerging exotic phenomena such as quasiparticle renormalization, strong correlations, and collective excitations [3–11].

For light-matter interactions, the QPT-simplified Lorentz oscillator model [12] has long been used as a standard model to determine the dielectric function $\varepsilon(\omega)$,

$$\varepsilon = \varepsilon_\infty\left(1 + \frac{\omega_{LO}^2 - \omega_{TO}^2}{\omega_{TO}^2 - \omega^2 - i\omega\gamma}\right) \tag{1}$$

where $\varepsilon_\infty$ represents the dielectric function at the high-frequency limit. The transverse and longitudinal optical phonon frequencies, $\omega_{TO}$ and $\omega_{LO}$, are well-defined within the harmonic and virtual crystal approximations, which neglects anharmonicity and random distribution of isotopes, respectively. In the presence of weak anharmonicity and isotope disorder, phonon modes acquire a finite linewidth $\gamma$, which can be evaluated using QPT by treating the anharmonicity and isotope disorder as perturbations to the harmonic virtual-crystal Hamiltonian. However, in the strong-perturbation limit, the picture of well-defined optical phonon modes ceases to apply. The coherently mixed vibrational modes in strong scattering regime complex light-matter interactions, requiring theoretical models beyond QPT.

Boron arsenide (BAs) has emerged as a standout semiconductor for thermal management because it combines ultrahigh lattice thermal conductivity [13–16] with ultrahigh carrier mobility [17,18], making it attractive for heat spreading and carrier transport in next-generation electronics. The integration of BAs into GaN-on-BAs devices has demonstrated an 8-fold enhancement on interfacial thermal conductance than GaN-on-diamond [19]. While *ab initio* theory has long predicted exceptional thermal transport in BAs [20,21], recent measurements report thermal conductivities that can exceed current *ab initio* theory expectations, leaving the microscopic origin of this discrepancy unresolved [22,23]. A key clue lies in the anomalous isotope physics of BAs. In naturally abundant samples, Raman spectroscopy exhibits a two-mode behavior that indicates coherent mixing between vibrations associated with different isotopic masses, rather than a simple broadening around a single virtual-crystal mode [24–26]. These observations suggest that isotope effects in BAs involve physics beyond the standard perturbative picture, motivating a nonperturbative framework for accurately capturing isotope effects in BAs.

*Contact author: huanwu@asu.edu

Here, we reveal that coherently mixed isotope vibrations in BAs render the light-matter interactions fundamentally distinct from the predictions of QPT, enabled by significant vibrational frequency splitting due to the large mass difference between boron isotopes. A nonperturbative approach is developed to capture the coupling between isotope vibrations and the resulting dielectric response. A mode-resolved criterion is introduced to quantify the breakdown of the perturbative description in BAs. We further explore the isotope-disorder effect on coupled surface phonon polaritons and near-field radiative heat transfer, and show that varying isotope composition enables a two-fold tuning of the radiative heat flux. The results demonstrate how nonperturbative isotope effects govern light-matter interactions in regimes inaccessible to conventional perturbative descriptions.

## II. THEORETICAL BACKGROUND

In general, the dynamic response of bound ions in a lattice to an oscillating external electromagnetic (EM) field involves atomic displacements driven by the electric field and counteracted by interatomic restoring forces, giving rise to frequency-dependent polarization [12,27]. The dielectric function characterizes how strongly this polarization responds to the applied field. When the frequency of the external field approaches the intrinsic phonon frequencies (without including long-range dipole-dipole interactions), the EM wave resonates with the lattice vibrations, causing the bound ions to oscillate strongly and leading to a sharp change in the dielectric function. Since phonon frequencies depend on atomic mass, the presence of multiple stable isotopes in an element leads to a distribution of vibrational frequencies. Within the virtual crystal approximation (VCA), an average mass is assigned to all atoms of the same element, which preserves translational symmetry and yields well-defined phonon modes. But in real materials, phonon frequency varies with the isotope mass, leading to the broadening of the phonon mode energy. When the isotope mass differences are small, the resulting finite linewidth can be evaluated using QPT. However, if the mass differences are large enough to cause substantial differences in phonon energies, the perturbative approach breaks down. Recent studies have shown that QPT fails to accurately describe the optical phonon modes in BAs due to the large mass difference between the two boron isotopes, $^{10}$B and $^{11}$B [25,26]. The random distribution of boron isotopes breaks the translational symmetry, and the interactions between isotopes give rise to coherent mixing of their vibrational modes [26]. Consequently, the light-matter interactions in BAs are complicated due to the large isotopic mass contrast, which results in the splitting and coupling of vibrational modes associated with different isotopes.

## III. METHOD

To elucidate the complex light-matter interactions in BAs, we start from the origin of Lorentz oscillator model, where bound ions oscillate under the external electric field and the interatomic restoring forces [27],

$$m_j \ddot{u}_{j\alpha} + m_j \gamma_j \dot{u}_{j\alpha} + \sum_{k\beta} \Phi_{j\alpha,k\beta} u_{k\beta} = \sum_\beta Z_{j,\alpha\beta} E_\beta(t) \tag{2}$$

where $\alpha$ and $\beta$ denotes the directions, and $u_{j\alpha}$ is the displacement of atom $j$ in the crystal driven by external electric field $E_\alpha$. $\Phi_{j\alpha,k\beta}$ is the harmonic interatomic force constant matrix. $m_j$, $\gamma_j$, and $Z_{j,\alpha\beta}$ are mass, damping coefficient, and born effective charge tensor of atom $j$, respectively. Note that lattice translational symmetry has not yet been introduced; therefore, Eq. (2) should apply to the entire crystal rather than the primitive cell. For practicality, we employ a supercell for the computation.

Transform Eq. (2) from real space to normal coordinates $Q_s$,

$$\ddot{Q}_s + \gamma_s \dot{Q}_s + \omega_s^2 Q_s = \sum_\alpha Z^*_{s,\alpha} E_\alpha \tag{3}$$

where $\omega_s$ is the vibrational frequency of mode $s$. $Z^*_{s,\alpha}$ is the mode effective charge,

$$Z^*_{s,\alpha} = \sum_{j\beta} Z_{j,\alpha\beta} \frac{e_{js,\beta}}{\sqrt{m_j}} \tag{4}$$

where $e_{js,\beta}$ is the $\beta$ component of the vibrational eigenvector of atom $j$ in mode $s$, determined by diagonalizing the dynamical matrix $D$ of the supercell,

$$D\mathbf{e}_s = \omega_s^2 \mathbf{e}_s \tag{5}$$

An applied electric field of frequency $\omega$ produces a normal mode amplitude,

$$Q_s = \frac{\sum_\alpha Z^*_{s,\alpha} E_\alpha}{\omega_s^2 - \omega^2 - i\omega\gamma_s} \tag{6}$$

The applied electric field can induce a polarization, $P_\alpha = \epsilon_0 \sum_\beta (\varepsilon_{\alpha\beta} - \delta_{\alpha\beta}) E_\beta$, where $\epsilon_0$ is the vacuum permittivity and $\delta_{\alpha\beta}$ is Kronecker delta function. Polarization is formed due to electronic response ($\boldsymbol{P}_{el}$) and ionic dipoles ($\boldsymbol{P}_{ion}$), such that $\boldsymbol{P} = \boldsymbol{P}_{el} + \boldsymbol{P}_{ion}$. The electrons response instantaneously on the timescale of lattice vibrations, hence $P_{el,\alpha} =$ $\epsilon_0 \sum_\beta (\varepsilon_{\infty,\alpha\beta} - \delta_{\alpha\beta}) E_\beta$ , where $\varepsilon_{\infty,\alpha\beta}$ is the high-frequency dielectric constant tensor. The ionic contribution can be written as the sum of the dipole moment, $P_{ion,\alpha} = 1/V \sum_s Z^*_{s,\alpha} Q_s$ , where $V$ is the volume of the crystal. Combining with Eq. (6), we determine the dielectric function,

$$\varepsilon_{\alpha\beta} = \varepsilon_{\infty,\alpha\beta} + \frac{1}{\epsilon_0 V} \sum_s \frac{Z^*_{s,\alpha} Z^*_{s,\beta}}{\omega_s^2 - \omega^2 - i\omega\gamma_s} \qquad (7)$$

where the summation is over all optical modes of the crystal. Since the formulation is not restricted to the primitive cell, the system contains $3(N-1)$ optical modes, where $N$ is the number of atoms in the supercell. $\gamma_s$ represents the linewidth from anharmonicity excluding mass-disorder effect.

For BAs, we take $\gamma_s$ from *ab initio* calculations considering both three and four phonon scatterings and excluding isotope scatterings [28]. Isotope disorder is treated by a nonperturbative approach, whereas anharmonicity is treated by QPT through $\gamma_s$.

In face-centered cubic BAs, $Z_{j,\alpha\beta}$ and $\varepsilon_{\infty,\alpha\beta}$ are diagonal and isotropic, with $Z_{\mathrm{As}} = +0.46$ , $Z_{\mathrm{B}} = -0.46$, and $\varepsilon_\infty = 9.83$ [29]. The lattice constant of BAs is measured to be 4.78 Å [13,30,31].

A supercell exact diagonalization approach is employed to nonperturbatively derive the eigenmodes of mass-disordered lattice beyond the VCA. Different from QPT, exact diagonalization solves the full eigenproblem in a finite Hilbert space, without taking the assumption of decomposing the system into non-interacting and perturbative components [32,33]. In principle, the eigenmodes of the entire crystal with randomly distributed isotopes can explicitly capture the coherent mixing of vibrations within the isotope-disordered system. However, computing the eigenmodes of the entire mass-disordered crystal requires prohibitively large computational resources. Hence, we construct a 512-atom supercell with periodic boundaries, and randomly assign an isotope type to each atomic site within the supercell. Each mass configuration represents a local snapshot of the mass-disordered system. By diagonalizing the dynamical matrix of the mass-disordered supercell, we obtain the exact eigenfrequencies $\omega_s$ and eigenvectors $\mathbf{e}_s$ of the mass-disordered system, which are used to derive the frequency-dependent dielectric function. We randomly generated 10000 mass configurations, and derived the averaged dielectric function from all individual configurations. 10000 mass configurations reach convergence with relative standard error for $|\varepsilon|$ less than 0.32% [34]. The supercell size is sufficiently large to capture the coherent mixing of boron isotope vibrations, as evidenced by the stable spectral features, where the primary (~700 cm$^{-1}$) and secondary (~720 cm$^{-1}$) resonance frequencies shift by merely 0.57 cm$^{-1}$ and 1.15 cm$^{-1}$, respectively, when the supercell is enlarged to 1000 atoms [34].

The interatomic force constants were derived, as in Ref. [26], by fitting the displacement-force datasets using ALAMODE [35]. The displacement-force datasets were obtained by computing interatomic forces for each irreducible displacement configuration via density functional theory [36,37] using Quantum ESPRESSO [38,39]. We used projector-augmented wave pseudopotentials [40]. Non-local exchange-correlation functional 'vdw-df2-c09' is employed for accurate long-range interatomic interactions [41]. The convergence threshold for self-consistency is $10^{-12}$. The kinetic energy cutoff for electronic wavefunctions is 100 Ry. Only $\Gamma$ point is considered for the $\boldsymbol{k}$-point grid of the supercell. As interatomic forces are governed solely by the electronic structure and independent of isotopic mass, the interatomic force constants are kept constant across all mass configurations.

The supercell exact diagonalization approach has been validated for determining the eigenmodes of isotope-disordered BAs by the excellent agreement between the measured and theoretical Raman spectra [26].

The key difference between our developed nonperturbative approach and QPT is that QPT relies on VCA, which replaces isotopic masses with their concentration-weighted averages. Within the QPT formalism, the eigenmodes are derived from virtual crystal, and the effect of isotope disorder is introduced through the linewidth determined by Fermi's golden rule [42], without altering the eigenmodes. In contrast, our nonperturbative approach determines the mode effective charge $Z^*_{s,\alpha}$ from the exact eigenmodes of the mass-disordered supercell, without decomposing the Hamiltonian into a non-interacting virtual-crystal component and a perturbative isotope-disorder component. The dielectric function spectrum is collectively contributed by all eigenmodes with optically active non-zero $Z^*_{s,\alpha}$ , each at its distinct frequency. Applying VCA to BAs supercell yields only three degenerate eigenmodes with non-zero $Z^*_{s,\alpha}$ due to translational symmetry, corresponding to the Γ-point optical mode of the virtual-crystal primitive cell, meaning that the supercell exact diagonalization approach inherently captures isotope-disorder effects across both perturbative and nonperturbative regimes.

## IV. RESULTS AND DISCUSSION

### A. Breakdown of QPT and emergence of the nonperturbative dielectric response

Figure 1 presents the dielectric function of naturally abundant BAs at 300 K calculated by nonperturbative approach, contrasted with the results from QPT. The QPT derives the dielectric function near the optical phonon frequency with features similar to those of most common polar crystalline materials. In Fig. 1(a), as the frequency increases, the real part of the dielectric function ($\varepsilon'$), representing the in-phase polarization response, rises sharply when the driving frequency approaches the phonon resonance, peaking when the external field resonates with the lattice vibrations. Beyond the resonant frequency, $\varepsilon'$ drops steeply and can even become negative, as the lattice vibrations lag behind the oscillating external field and respond out of phase. In Fig. 1(b), the imaginary part of the dielectric function ( $\varepsilon''$ ), associated with absorption, shows a pronounced peak at the intrinsic phonon frequency, where resonance strongly enhances the energy absorption by phonons.

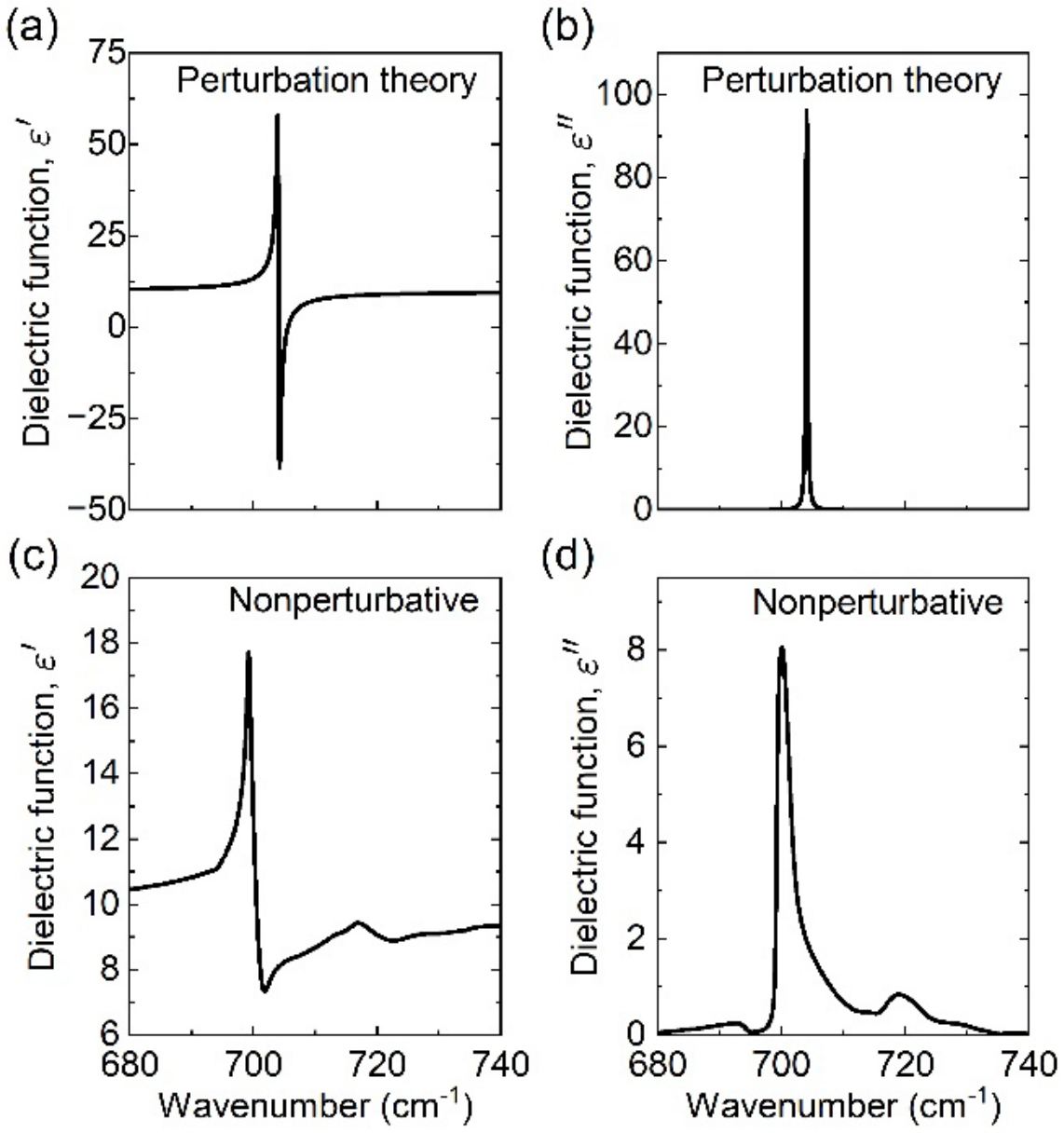

FIG. 1. Dielectric function $\varepsilon = \varepsilon' + i\varepsilon''$ of naturally abundant boron arsenide. The (a) real and (b) imaginary parts calculated within QPT differ fundamentally from the nonperturbative results shown in (c) and (d), demonstrating the failure of QPT in boron arsenide.

In contrast, the nonperturbative approach uncovers characteristics in dielectric function distinct from the results of QPT. In Fig. 1(c), the detailed spectral profile of $\varepsilon'$ exhibits many fine features, characterized by a dominant peak at 699.3 cm$^{-1}$, a valley at 701.8 cm$^{-1}$, a small bump at 716.9 cm$^{-1}$, and a shallow dip at 722.7 cm$^{-1}$. Unlike most materials that exhibit well-defined intrinsic phonon frequencies, the large mass difference between $^{10}$B and $^{11}$B causes their vibrations to occur at very different frequencies, preventing the formation of a single collective mode near the virtual-crystal frequency. The isotope disorder produces coherently mixed vibrational modes between $^{10}$B and $^{11}$B, forming a broad spectrum in dielectric function. The dominant peak and valley are primarily associated with $^{10}$B, and the small bump and shallow dip are primarily associated with $^{11}$B. Different from the QPT, the $\varepsilon'$ at the valley or dip no longer falls below zero due to the positive compensation from the other isotope, leading the vibrations at the valley frequencies to no longer be out of phase with the external field. In Fig. 1(d), the $\varepsilon''$ shows one major peak at 700.1 cm$^{-1}$ and a bump at 718.9 cm$^{-1}$, primarily associated with the vibrational modes of $^{11}$B and $^{10}$B, respectively. The striking differences between the QPT and nonperturbative results demonstrate the failure of the QPT in determining the dielectric function of BAs.

The QPT framework cannot capture several key consequences of isotope disorder in BAs. First, mass disorder breaks lattice translational symmetry, effectively folding the Brillouin zone and projecting phonon modes with nonzero wavevectors to the Γ-point [26]. However, the VCA employed in QPT artificially restores the translational symmetry, thereby eliminating disorder-induced zone-folding and rendering the corresponding modes optically inactive. Second, $^{10}$B and $^{11}$B have markedly different vibrational frequencies, which leads to significant vibrational mode splitting [26]. By treating all boron as identical virtual atoms, the mean-field description removes the distinction between $^{10}$B and $^{11}$B, thereby merging their vibrational modes. Third, $^{10}$B and $^{11}$B interact through interatomic forces and jointly form normal modes, giving rise to coherent mixing between their vibrations, which cannot be captured by QPT and necessitate nonperturbative treatment.

### B. Temperature robustness of the nonperturbative dielectric features

We next examine whether the nonperturbative dielectric response in naturally abundant BAs remains robust against thermal broadening. As the anharmonic scattering usually become stronger with increasing temperature, the dielectric function typically broadens at elevated temperatures due to the enhanced $\gamma_s$. Nevertheless, the temperature-dependent dielectric function, shown in Fig. 2, reveals that the fine spectral features are sustained at elevated temperatures up to 1000 K. For $\varepsilon'$, the peak at 716.9 cm$^{-1}$ persists up to 1000 K, despite broadening. For $\varepsilon''$, the peak at 718.9 cm$^{-1}$ remains distinct up to 800 K, gradually evolving into a shoulder as the temperature reaches 1000 K. The persistence of the spectral features at elevated temperatures is attributable to the substantial energy splitting between $^{10}$B and $^{11}$B vibrational modes.

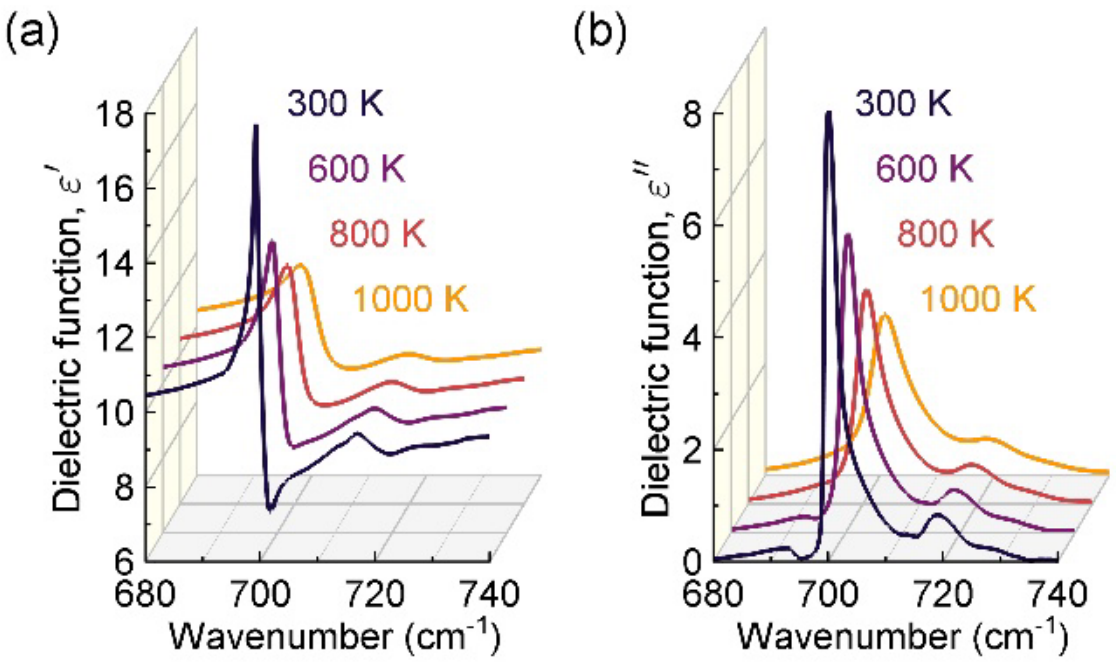

FIG. 2. (a) Real and (b) imaginary part of the dielectric function calculated for different temperatures.

### C. Mode-resolved criterion for the breakdown of the perturbative description

To quantitatively justify whether the isotope disorder is out of the perturbative regime, we introduce a mode-resolved criterion based on Green's function. The dynamical matrix of a mass-disordered supercell can be separated as

$$D = D_0 + V \tag{8}$$

where $D_0$ is the dynamical matrix of virtual-crystal supercell.

The retarded Green's function is

$$G = [(\omega^2 + i0^+)I - D]^{-1} \tag{9}$$

and the corresponding virtual crystal Green's function is,

$$G_0 = [(\omega^2 + i0^+)I - D_0]^{-1} \tag{10}$$

The Dyson's equation, $G = G_0 + G_0 V G$, can be written as

$$G = [I - G_0 V]^{-1} G_0 = \sum_{n=0}^{\infty} [G_0 V]^n \, G_0 \tag{11}$$

provided that this Neumann series converges. This yields a natural perturbativity indicator: the expansion converges when the dimensionless operator $G_0 V$ is small in norm.

For numerical stability we use the equivalent resolvent form for $G_0$,

$$R_0(z) = [D_0 - zI]^{-1}, \text{ where } z = \omega^2 + \eta i \tag{12}$$

The broadening $\eta$ prevents divergences at poles. In practice, we set $\eta = 2\omega\gamma$, where $\gamma$ is the anharmonic linewidth. With this formula, a sufficient convergence criterion for the Neumann series becomes $\|R_0 V\| \ll 1$.

According to Eq. (7), only modes with large $|Z^*_{s,\alpha}|$ contribute significantly to light-matter interactions, so we restrict the perturbative analysis to the subset of optically active modes $\mathcal{S} \equiv \{s: |Z^*_{s,\alpha}|/|Z_{\alpha\alpha}| > 0.005\}$, i.e., modes with effective charge $|Z^*_{s,\alpha}|$ exceeds 0.5% of the corresponding Born effective charge $|Z_{\alpha\alpha}|$. The orthogonal projector onto the subspace spanned by $\{\mathbf{e}_s\}_{s\in\mathcal{S}}$ is $P = \sum_{s\in\mathcal{S}} \mathbf{e}_s \mathbf{e}_s^\dagger$, which is Hermitian and idempotent.

We then define the projected resolvent-mediated mixing operator

$$\mathcal{R}(z) \equiv P R_0(z) V P \tag{13}$$

and the scalar indicator

$$r(z) = \|\mathcal{R}(z)\|_2^2 \tag{14}$$

where $r(z)$ is the spectral norm square of $\mathcal{R}(z)$, measuring the strength of disorder-induced multiple scattering within the optically active subspace at frequency $\omega$. $r(z) > 1$ indicates the breakdown of perturbative description for the optical response at $\omega$.

We calculated $r(z)$ for common polar materials prototyped by GaAs and 3C-SiC. The interatomic force constants are derived using projector-augmented wave pseudopotentials under local density approximation [43]. The kinetic energy cutoffs for electronic wavefunctions are 80 and 120 Ry for GaAs and 3C-SiC, respectively. The supercells contain 216 atoms, and the $\boldsymbol{k}$-point grids are 2×2×2. The anharmonic linewidths for GaAs and 2C-SiC were extracted from Ref. [44] through the differences between the total linewidths at 300 K and 0 K, since isotope scattering dominates at 0 K and is independent of temperature. The high-frequency dielectric constants and Born effective charges for GaAs and 3C-SiC are taken from Ref. [45] and [46], respectively.

Figure 3 shows the results for $r(z)$, which is averaged over an ensemble of random mass configurations. For GaAs, $r < 0.0008$, and for 3C-SiC, $r < 0.05$, confirming those systems reside in the perturbative regime. For BAs, however, $r$ reaches 340 at 300 K, indicating strong isotope disorder beyond the perturbative regime. Even at 1000 K, BAs still exhibits a large $r$ with a peak value of 2.9, consistent with the persistence of mode-splitting features in dielectric function up to 1000 K (Fig. 2).

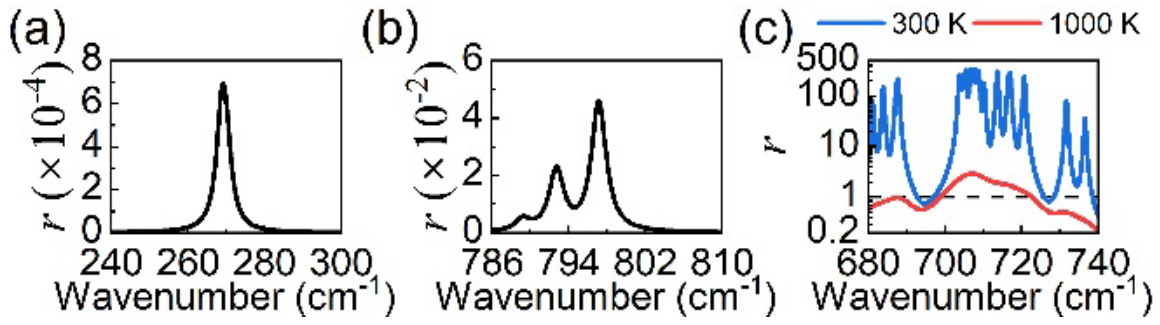

FIG. 3. The indicator $r$ distinguishing perturbative and nonperturbative regimes for (a) GaAs, (b) 3C-SiC, and (c) BAs.

### D. Composition-driven perturbative-to-nonperturbative crossover

Since natural-abundance BAs lies deep in the nonperturbative isotope-disorder regime, we trace the evolution toward the perturbative limit as the isotope composition varies. To characterize the perturbative-to-nonperturbative crossover of isotope disorder, we calculate the dielectric function for different isotope compositions using supercell exact diagonalization approach, as shown in Fig. 4.

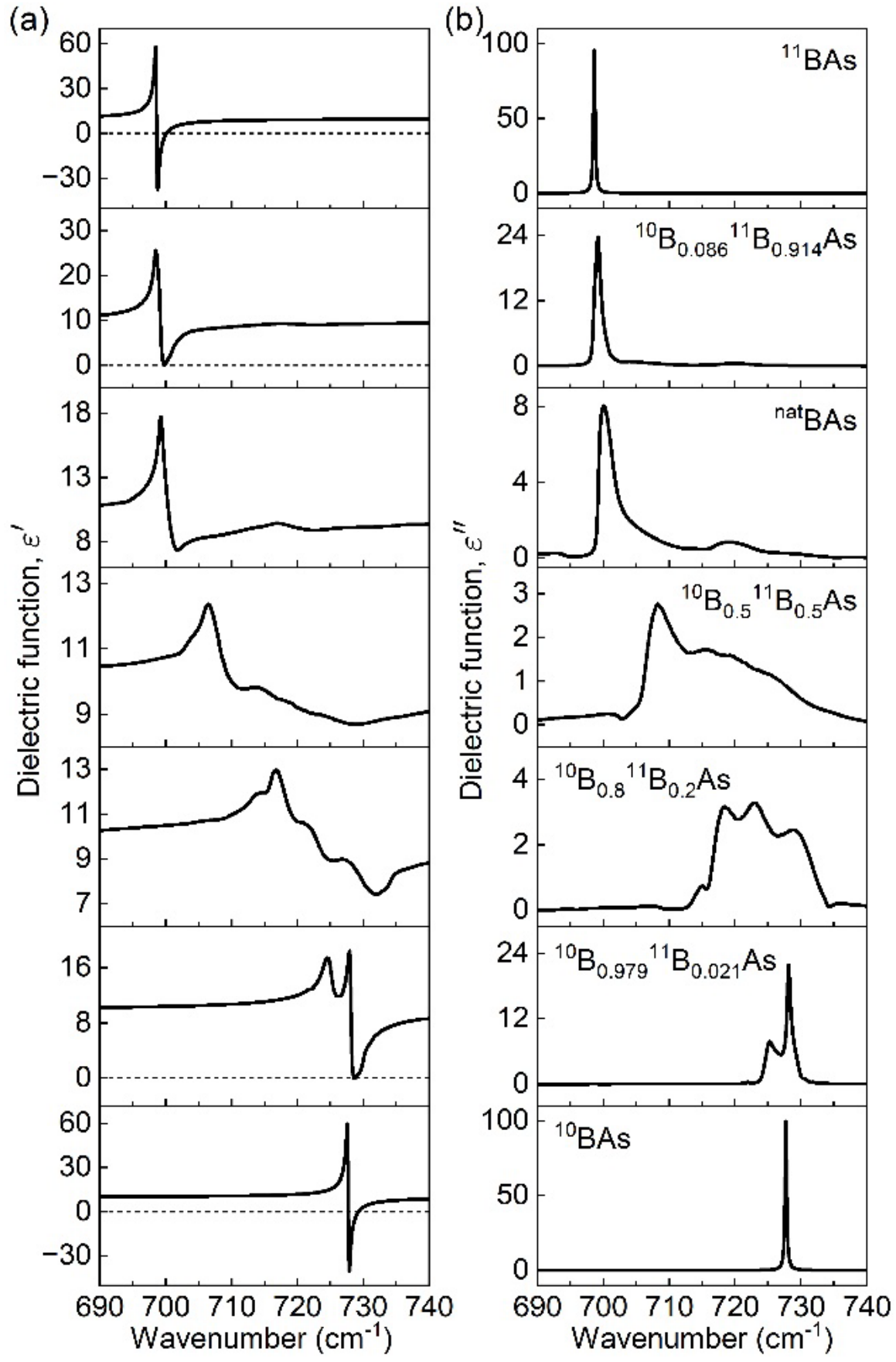

FIG. 4. (a) Real and (b) imaginary part of the dielectric function calculated for varying isotope composition. $^{10}$B concentration is 19.9% for naturally occurring isotope composition.

Figure 4(a) shows that, as the isotope composition varies, $\varepsilon'(\omega)$ evolves from having negative values in the perturbative regime to remaining positive across the entire spectrum. As the composition approaches either pure $^{10}$B or $^{11}$B, the system tends toward the perturbative limit, in which a single isotope dominates to produce only one resonant frequency. Slightly above the resonant frequency, the polarization lags the driving field, causing $\varepsilon'$ to drop deeply into a pronounced negative minimum. At the natural abundance of 19.9% $^{10}$B, $\varepsilon'(\omega)$ is positive across spectrum. This occurs because the total dielectric response is a superposition of contributions from all isotopes ($^{10}$B and $^{11}$B). At frequencies where the $\varepsilon'(\omega)$ contributed by one isotope would typically become negative, the $\varepsilon'(\omega)$ contributed by the other isotope is still positive, hence the total $\varepsilon'(\omega)$ contributed by all isotopes remains positive. The compensation between individual resonance tails of $^{10}$B and $^{11}$B makes the sum positive across spectrum. When the $^{10}$B concentration reaches either 8.6% or 97.9%, the minimum of $\varepsilon'(\omega)$ touches zero, defining the critical boundary between the regime where $\varepsilon'(\omega)$ can be negative and where $\varepsilon'(\omega)$ is positive across spectrum.

Figure 4(b) demonstrates $\varepsilon''(\omega)$ evolve with isotope composition. $\varepsilon''(\omega)$ is associated with the EM wave absorption through the interactions with atomic vibrations. Instead of a single peak shifting continuously as predicted by VCA, the nonperturbative results reveal the simultaneous presence of multiple resonance features associated with vibrations dominated $^{10}$B and $^{11}$B, respectively. As the $^{10}$B concentration increases from 0 to 19.9% (natural abundance), the major resonant peak gradually blueshifts and weakens, while a small bump associated with $^{10}$B vibrations emerges near 720 cm$^{-1}$. In the intermediate regime (e.g., 50% and 80% $^{10}$B), multiple vibrational modes overlap to form a broadened, multi-peaked profile. With further enrichment of $^{10}$B, the $^{10}$B-dominated peak at higher frequency becomes stronger than the $^{11}$B-dominated peak at lower frequency. When the system approaches an almost pure isotope limit, the disorder is strongly reduced, and the dielectric response recovers to single resonance behavior.

The evolution of $\varepsilon''(\omega)$ resembles the variations of the eigenmodes of the mass-disordered system. At low $^{10}$B concentration, the lattice dynamics are primarily $^{11}$B-like, and $^{10}$B acts as a dilute impurity perturbation that widens the $^{11}$B resonance peak. As the $^{10}$B fraction increases, $^{10}$B hybridizes with $^{11}$B vibrations and form extended resonant modes, leading to substantial mode mixing and spectral broadening. In the $^{10}$B-rich regime, the dominant eigenmodes become $^{10}$B-like, while $^{11}$B behaves as a heavy impurity that contributes a weaker low-frequency resonance and broadens the $^{10}$B resonance peak. Overall, the complex dielectric function spectra reveal signatures of the isotope-induced evolution of eigenmodes and their hybridization, which can be characterized experimentally by spectroscopic ellipsometry [47].

### E. Isotope effect on radiative heat transfer across separation regimes

With the nonperturbative dielectric function of isotope-disordered BAs determined, we then investigate how isotope disorder influences radiative heat transfer. The total radiative heat flux $Q$ is the integration over all spectral contributions,

$$Q = \int_0^\infty q(\omega)\mathrm{d}\omega \tag{15}$$

where $q(\omega)$ is spectral radiative heat flux per unit area [12,48],

$$q(\omega) = \frac{\hbar\omega[n(\omega,T_1) - n(\omega,T_2)]}{4\pi}\int_0^\infty \xi(\omega,\beta)\beta\mathrm{d}\beta \tag{16}$$

where $n(\omega, T)$ is the Bose-Einstein distribution function. The heat flux is driven by the temperature difference between the two bulks, $T_1$ and $T_2$. $\xi(\omega, \beta)$ is the energy transmission coefficient accounting for both transverse electric ($s$-polarized) and transverse magnetic ($p$-polarized) waves, and it is expressed separately for propagating and evanescent waves [12,48]. For propagating waves ($\beta < k_0$),

$$\xi = \sum_{m=s,p} \frac{\left(1-|r_{01}^m|^2\right)\left(1-|r_{02}^m|^2\right)}{\left|1-r_{01}^m r_{02}^m \exp(2ik_{z0}d)\right|^2} \tag{17}$$

For evanescent waves ($\beta > k_0$),

$$\xi = \sum_{m=s,p} \frac{4\mathrm{Im}(r_{01}^m)\mathrm{Im}(r_{02}^m)\exp[-2\mathrm{Im}(k_{z0}d)]}{\left|1-r_{01}^m r_{02}^m \exp[-2\mathrm{Im}(k_{z0}d)]\right|^2} \tag{18}$$

Here, $s$ and $p$ stand for $s$-polarized and $p$-polarized waves, respectively. $r_{0j}$ is the Fresnel coefficient at the interface between vacuum (0) and bulk ($j$). For $s$-polarized waves,

$$r_{0j}^s = \frac{k_{z0} - k_{zj}}{k_{z0} + k_{zj}} \tag{19}$$

For $p$-polarized waves,

$$r_{0j}^p = \frac{\varepsilon_j k_{z0} - k_{zj}}{\varepsilon_j k_{z0} + k_{zj}} \tag{20}$$

To systematically understand the influence of isotope disorder on radiative heat transfer across different separation regimes, we calculated the radiative heat flux between two BAs bulks as a function of gap distance ($d$) for several isotope compositions, as shown in Fig. 5.

At sufficiently large $d$, evanescent wave transmission becomes exponentially small, as described by Eq. (18). In this regime, radiative heat transfer is dominated by propagating waves, placing the system firmly in the far-field regime. In far-field regime, the radiative heat flux is independent from $d$, as shown in Fig. 5(a) at large $d$ limit.

When $d$ falls below the thermal wavelength $\lambda_T = \hbar c/k_B T$ (7.6 μm at 300 K), which characterizes the dominant wavelength of thermally excited photons [49], radiation enters the near-field regime where evanescent waves dominate heat transfer. Evanescent waves are exponentially decaying EM fields bound to surfaces. When two surfaces are separated by less than the evanescent decay length, the fields can tunnel across the gap, significantly enhancing the radiative heat flux. The near-field radiative heat flux depends strongly on $d$, as shown in Fig. 5(a), because reducing the gap enhances evanescent tunneling.

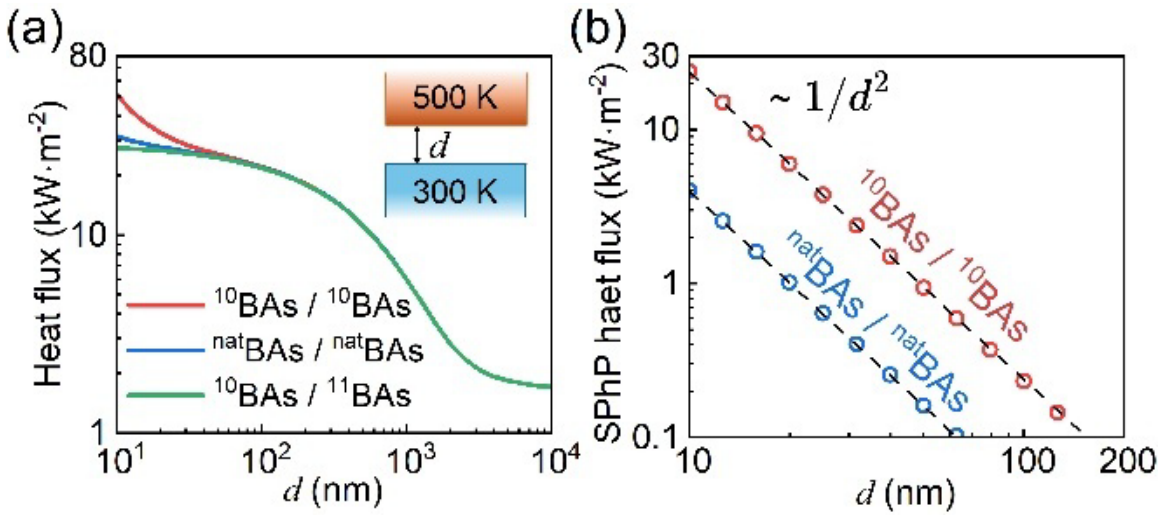

FIG. 5. (a) Total and (b) SPhP-mediated near-field radiative heat flux between BAs bulks under varying gap distance $d$ for different isotope compositions. The symbols in (b) denote the calculated results, and the dashed lines guide the $1/d^2$ scaling. The $^{10}$BAs / $^{11}$BAs curve is not shown in (b) because it is near zero.

The near-field radiation can be divided into two regimes depending on whether surface phonon polariton (SPhP) causes a significant enhancement on heat transfer. SPhP emerges through the resonance between the bound ion vibration and the EM wave near the surface. At sufficiently small $d$, SPhPs on the two surfaces can interact to form coupled-SPhP (c-SPhP), enabling resonant energy exchange that substantially enhance the near-field radiative heat flux [50]. By contrast, when $d$ is relatively large within the near-field regime, the inter-surface coupling of SPhPs remains weak, and the heat transfer is dominated by non-resonant evanescent waves associated with frustrated total internal reflection.

When $d$ is reduced below 100 nm, near-field radiation enters the SPhP-mediated regime, where the radiative heat flux demonstrates a pronounced dependence on isotope compositions, as shown in Fig. 5(a). To quantitatively investigate the isotope-disorder effect on SPhP-mediated near-field radiation, we isolate the c-SPhP contribution by focusing on the $p$-polarized evanescent waves, since SPhP are supported exclusively by $p$-polarized evanescent waves. Specifically, we compute the heat flux carried by $p$-polarized evanescent waves using the full dielectric function $\varepsilon(\omega)$, and subtract the non-resonant frustrated contribution estimated by replacing $\varepsilon(\omega)$ with the high-frequency dielectric constant $\varepsilon_\infty$.

Figure 5(b) shows the isotope composition strongly tunes the SPhP-mediated radiative heat flux. $^{10}$BAs / $^{10}$BAs shows the highest SPhP-mediated heat flux, because the pronounced peak in $^{10}$BAs dielectric function indicates weak damping in the resonance between ionic vibration and surface EM wave, which promotes the inter-surface resonance of SPhPs. $^{nat}$BAs / $^{nat}$BAs shows a relatively lower SPhP-mediated heat flux, as isotope disorder smears out the dielectric response and weakens the SPhP coupling across the gap. $^{10}$BAs / $^{11}$BAs is not shown in Fig. 5(b) due to the nearly vanishing SPhP-mediated radiative heat flux. Though both $^{10}$BAs and $^{11}$BAs have a pronounced peak in dielectric function, $^{10}$BAs and $^{11}$BAs support SPhPs at very different optical phonon frequencies, so the SPhPs on the two sides cannot satisfy the resonance condition for efficient energy exchange.

Moreover, Fig. 5(b) shows that the calculated SPhP-mediated radiative heat flux scales as $\sim 1/d^2$, as $\xi \sim 1/d^2$ in the limit of small $d$ according to Eq. (18).

### F. Coupled surface phonon polaritons in the nonperturbative isotope-disorder regime

To clarify the microscopic origin of the strong isotope-composition dependence of the SPhP-mediated radiative heat flux, we next investigate the c-SPhP dispersion relation and the energy transmission coefficient. The dispersion relation of c-SPhP is the relationship between the in-plane wavevector $\beta$ and the wavenumber $\omega$, which can be determined from the solution of Maxwell equations on vacuum sandwiched between two bulks (Fig. 6(b) inset) [12,51],

$$\tanh(ik_{z0}d)\left(\frac{k_{z0}^2}{\varepsilon_0^2}+\frac{k_{z1}k_{z2}}{\varepsilon_1\varepsilon_2}\right)=\frac{k_{z0}}{\varepsilon_0}\left(\frac{k_{z1}}{\varepsilon_1}+\frac{k_{z2}}{\varepsilon_2}\right) \quad (21)$$

where the subscripts 1 and 2 refer to the two BAs bulks, and 0 denotes the vacuum region. The subscript $z$ indicates cross-plane component. $d$ is the distance of the vacuum gap. $\beta$ is identical in the two bulks and the gap due to the phase-matching boundary condition. The cross-plane wavevector $k_{zi}$ varies among the three media and obeys the relation $\varepsilon_i k_0^2 = k_{zi}^2 + \beta^2$, where $k_0 = \omega/c$ is the wavevector in vacuum and $c$ is the speed of light. For c-SPhP, the imaginary part of $k_{zi}$ is large due to the need of balancing $\beta$, causing a rapid cross-plane decay of evanescent wave and confinement near the gap.

Figure 6 (a) and (b) show the dispersion relation of c-SPhP in a 10-nm vacuum gap, derived using the dielectric functions from QPT and nonperturbative approaches, respectively. The dispersion relation by QPT exhibits a sharp peak at 705.4 cm$^{-1}$, signifying the strong resonance between EM wave and bound ion vibrations within a narrow frequency band. In contrast, by nonperturbative approach, the dispersion relation spans from ~690 cm$^{-1}$ to ~720 cm$^{-1}$, indicating the EM wave can resonate over a broader frequency range with vibrations of the bound ions with different masses. The large peak and small bump in the dispersion are associated primarily with $^{11}$B and $^{10}$B, respectively. The magnitude of the peak $\beta$ is around 40 times smaller than that from QPT, meaning for less polaritonic and more photon-like characteristics due to isotope disorder.

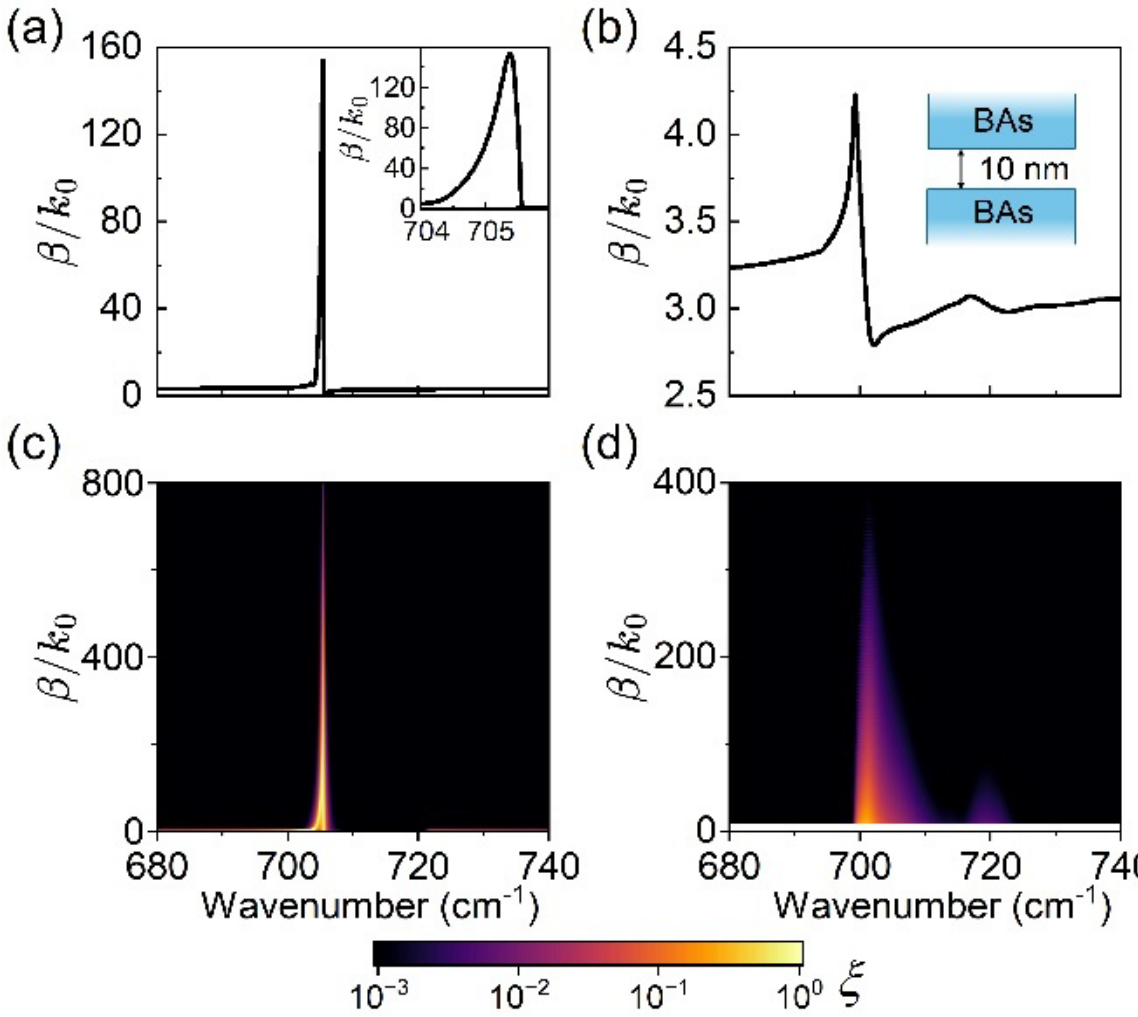

FIG. 6. Near-field radiative properties between naturally abundant BAs bulks. (a-b) Dispersion relation of c-SPhP calculated using (a) QPT and (b) nonperturbative approach, respectively. (c-d) Energy transmission coefficient $\xi$ calculated using (c) QPT and (d) nonperturbative approach, respectively. $k_0$ is the vacuum wavevector.

To quantitatively understand how isotope disorder influences the coupling of SPhPs across the gap, we compare the contour plots of the $\xi(\omega, \beta)$ derived through QPT and nonperturbative approach in Fig. 6 (c) and (d). Fundamentally distinct characteristics are found between the two approaches. Within QPT, high transmission coefficient occurs in a narrow frequency band. In contrast, the nonperturbative approach provides transmission over a broader frequency range, but with a substantially reduced transmission coefficient below 0.3. This reduction arises because strong isotope disorder in natural-abundance BAs disrupts the well-defined coherent lattice vibrations, smearing out the resonance that supports SPhP, and thereby diminishing energy transmission. The striking discrepancies highlight the breakdown of QPT in describing light-matter interactions in natural-abundance BAs. This comparison also implies that the transmission can be significantly enhanced once isotope disorder enters the perturbative regime. In isotopically pure BAs, the dielectric resonance becomes much sharper, which strengthens the inter-surface SPhP coupling and promotes energy transmission across the gap.

### G. Isotope tuning of SPhP-mediated radiative heat transfer

As isotope disorder strongly affects radiative heat flux in the SPhP-mediated regime, the heat transfer can be effectively tuned by isotope composition. To elucidate the tuning of radiative heat flux by isotope composition in the SPhP-mediated regime, we calculated the spectral radiative heat flux across a 10-nm vacuum gap for varying isotope composition, as shown in Fig. 7.

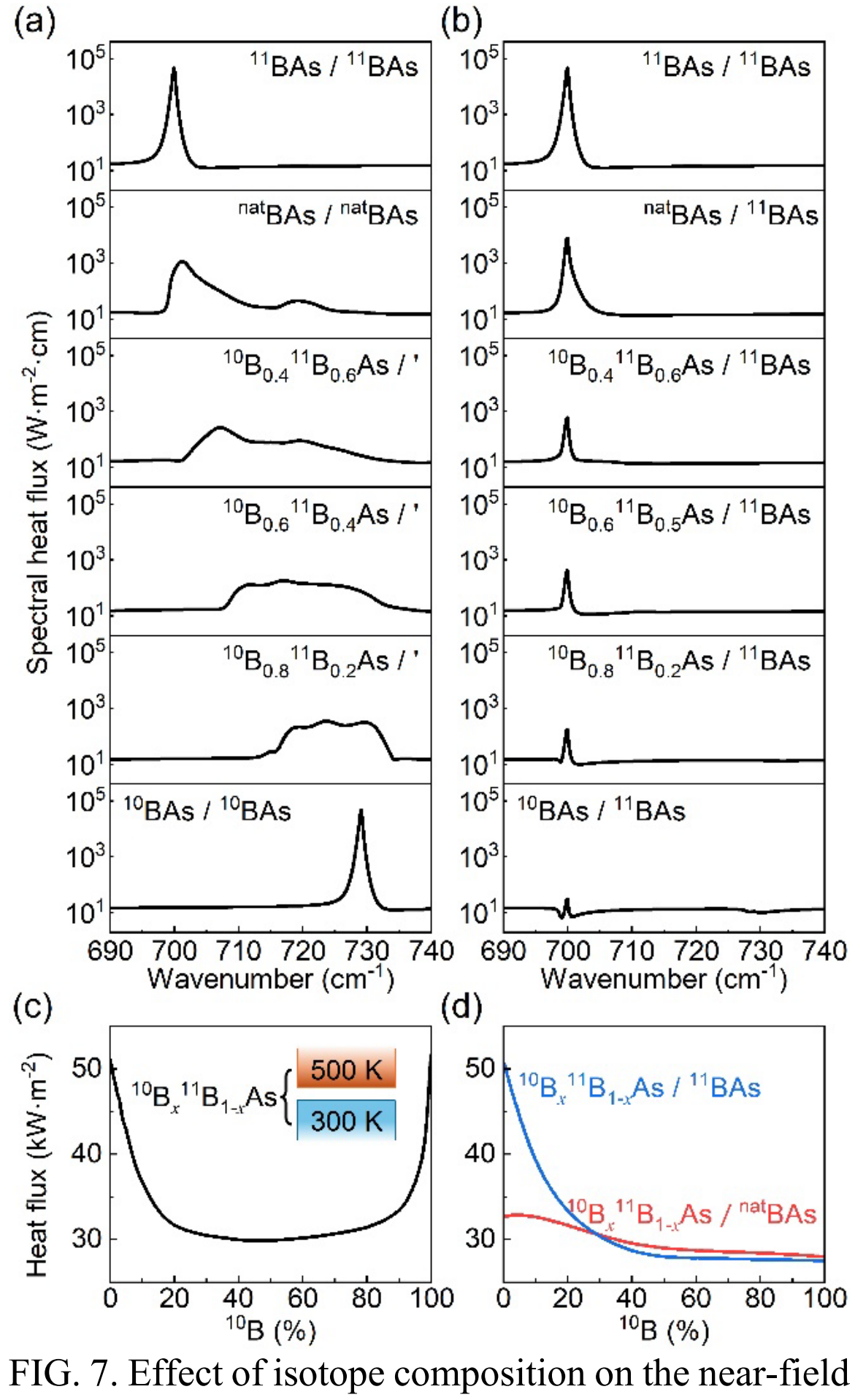

FIG. 7. Effect of isotope composition on the near-field radiative heat flux between BAs bulks separated by a 10 nm vacuum. The hot and cold sides are maintained at 500 K and 300 K, respectively. (a) Spectral heat flux for symmetric configurations with both sides at the same isotope composition. (b) Spectral heat flux for asymmetric configurations between $^{11}$BAs and BAs with a varying isotope composition. (c) Total heat flux varies with isotope composition for symmetric configurations with both sides at the same isotope composition. (d) Total heat flux for asymmetric configurations with one side under varying isotope composition and the other side under fixed isotope composition.

The radiative heat flux for a symmetric configuration with identical isotope compositions on both sides (inset of Fig. 7(c)) is shown in the left column of Fig. 7. The spectral heat flux is shown in Fig. 7(a), focusing on the spectral range of SPhP resonance. As the $^{10}$B concentration increases from the pure $^{11}$BAs (top panel) to $^{nat}$BAs, the peak at ~700 cm$^{-1}$ broadens and drops in magnitude by nearly two orders. At intermediate $^{10}$B concentrations (0.4~0.8), the spectral heat flux changes from a single resonance to a broad plateau. As the composition approaches pure $^{10}$BAs (bottom panel), a sharp resonance emerges at ~730 cm$^{-1}$. For isotopically pure BAs in the top and bottom panels of Fig. 7(a), a sharp peak is found in the spectral heat flux, as a well-defined intrinsic optical phonon frequency is established in the absence of mass-disorder. When the EM wave frequency approaches the intrinsic phonon frequency, strong resonance occurs, exciting c-SPhP in the vacuum gap. This resonance facilitates energy exchange mediated by the confined EM wave, making the heat exchange concentrate into a narrow spectral range near the well-defined phonon frequency, forming a sharp peak. With isotope disorder, the spatially varying isotope mass spreads the phonon frequencies, making the c-SPhP emerge in a wider frequency range. As the energy transmission is smeared out by isotope disorder, the spectral heat flux shows flattened peaks or plateau with lower magnitude for isotopically mixed BAs in the panels 2-5 of Fig. 7(a).

Figure 7(c) shows the total radiative heat flux as a function of isotope composition, exhibiting a U-shaped dependence, with the minimum at mixed isotope compositions reduced to almost half of the isotopically pure heat flux, reflecting the suppression of transmission by isotope disorder.

Figure 7(b) shows the spectral heat flux for an asymmetric configuration with $^{11}$BAs at the cold side and varying isotope composition at the hot side. Regardless of changes to the isotope composition on the hot side, the spectral heat flux consistently peaks at the cold side intrinsic phonon frequency, where lower temperatures reduce anharmonicity and sharpen SPhP resonance modes. As the $^{10}$B concentration on the hot side increases, the spectral heat flux decreases due to the growing mismatch between the intrinsic phonon frequencies of the two sides. This mismatch weakens the coupling between the SPhP modes at the two surfaces, thereby reducing the energy transmission. At maximum mismatch, i.e., for $^{10}$BAs/$^{11}$BAs in the bottom panel of Fig. 7(b), the spectral heat flux shows dips at certain wavenumbers that fall even below the nonresonant baseline. This occurs when the hot side supports a bound SPhP mode while the cold side acts as a nonresonant reflector. The opposite-phase response between the two surfaces leads to destructive interference in the transmission coefficient, producing a dip below the nonresonant baseline.

In Fig. 7(d), the total heat flux of $^{10}B_x{}^{11}B_{1-x}As/^{11}BAs$ decreases monotonically with increasing $^{10}B$ concentration due to the growing phonon mismatch between the two sides. When the cold side changes from $^{11}BAs$ to $^{nat}BAs$, the best phonon matching is expected for $^{nat}BAs/^{nat}BAs$, but the maximum total heat flux occurs at $^{10}B_{0.04}{}^{11}B_{0.96}As/^{nat}BAs$, resulting from a synthetic effect of isotope disorder and phonon matching.

By varying the isotope composition, a two-fold tuning of the heat flux can be achieved, showing the potential of isotope engineering in modulating the near-field radiation via SPhP resonance, providing a material-level mechanism to modulate near-field radiative heat transfer without altering geometry or surface chemistry. Such prediction can be measured through established nanoscale-gap platforms (e.g., sphere–plane or MEMS/AFM-based techniques) in the tens to hundreds of nanometers regime relevant to our calculations [50,52–58].

## V. SUMMARY

In summary, a nonperturbative approach based on the exact diagonalization of the isotope-disordered system is developed to capture the anomalous isotope effect on the light-matter interactions in BAs. The large boron isotope mass difference renders the light-matter interactions fundamentally distinct from the QPT behavior and fails the mean field approaches such as VCA. The coherently mixed isotope vibrations determine the profile of dielectric function in the nonperturbative regime, thereby dictating the c-SPhP properties and near-field radiation. The isotope disorder can disrupt the energy transmission by spreading the phonon spectrum or inducing a spectral mismatch between the two bulks. Isotope engineering achieves two-fold tuning of the radiative heat flux by modulating the SPhP resonance, without altering the geometry or surface chemistry. This work reveals the dominance of nonperturbative isotope physics in governing light-matter interactions where the conventional perturbative description fails. A unified theoretical framework that connects the perturbative and nonperturbative limits of isotope disorder is developed, enabling accurate predictions across weak to strong disorder regimes. We hope the theoretical predictions presented here will inspire innovative experiments for verification.

*Acknowledgements*—This work used Bridges-2 at Pittsburgh Supercomputing Center through allocation PHY230112 from the Advanced Cyberinfrastructure Coordination Ecosystem: Services & Support (ACCESS) program, which is supported by National Science Foundation grants #2138259, #2138286, #2138307, #2137603, and #2138296.

*Data availability*—The data that support the findings of this article are openly available [59].